\documentclass[epj]{svjour}
\usepackage{graphics}
\usepackage{graphicx}
\usepackage{placeins}
\usepackage[english]{babel}
\usepackage{blindtext}
\usepackage{caption}

\graphicspath{{figures/}} 
\begin{document}
\title{The SPEDE spectrometer}
\author{P. Papadakis\inst{1,2}
\and D.M. Cox\inst{1,2}
\and G.G. O'Neill\inst{3,}\thanks{Department of Physics, University of the Western Cape, P/B X17, Bellville, 7535, South Africa}
\and M.J.G. Borge\inst{4,}\thanks{On leave from Instituto de Estructura de la Materia, IEM, CSIC, Serrano 113bis, 28006-Madrid, Spain}
\and P.A. Butler\inst{3}
\and L.P. Gaffney\inst{4}
\and P.T. Greenlees\inst{1,2}
\and R.-D. Herzberg\inst{3}
\and A. Illana\inst{5}
\and D.T. Joss\inst{3}
\and J. Konki\inst{1,2}
\and T. Kr\"oll\inst{6}
\and J. Ojala\inst{1,2}
\and R.D. Page\inst{3}
\and P. Rahkila\inst{1,2}
\and K. Ranttila\inst{1}
\and J. Thornhill\inst{3}
\and J. Tuunanen\inst{1}
\and P. Van Duppen\inst{5}
\and N. Warr\inst{7}
\and J. Pakarinen\inst{1,2,}\thanks{Corresponding author, email: janne.pakarinen@jyu.fi}
}

\institute{University of Jyv\"askyl\"a, Department of Physics, P.O. Box 35, FI-40014 University of Jyv\"askyl\"a, Finland
\and Helsinki Institute of Physics, P.O. Box 64, FI-00014 University of Helsinki, Finland
\and Oliver Lodge Laboratory, University of Liverpool, Liverpool, L69 7ZE, United Kingdom\and EP Department, ISOLDE, CERN, CH-1211, Geneva, Switzerland
\and Instituut voor Kern-en Stralingsfysica, KU Leuven, B-3001 Leuven, Belgium
\and Institut f\"ur Kernphysik, Technische Universit\"at Darmstadt, D-64289 Darmstadt, Germany 
\and Institut f\"ur Kernphysik, Universit\"at zu K\"oln, Z\"ulpicher Stra{\ss}e 77, D-50937 K\"oln, Germany
}
\date{Received: 21.09.2017}
\abstract{
The electron spectrometer, {\sc spede}, has been developed and will be employed in conjunction with the Miniball spectrometer at the HIE-ISOLDE facility, CERN. {\sc spede} allows for direct measurement of internal conversion electrons emitted in-flight, without employing magnetic fields to transport or momentum filter the electrons. Together with the Miniball spectrometer, it enables simultaneous observation of $\gamma$ rays and conversion electrons in Coulomb-excitation experiments using radioactive ion beams.
} 

\maketitle
\section{Introduction}
\label{intro}
In-beam spectroscopic methods have extensively been employed to uncover the governing forces in atomic nuclei. The outcome of the development of detector materials and detection techniques has culminated in large $\gamma$-ray detector arrays and various electron spectrometers (see e.g. References \cite{Akkoyun2012AGATAArray,Eberth2008FromDetectors,VanKlinken1975Mini-orangeElectrons,Butler1996} and references therein). In order to establish a complete picture of de-excitation processes in a single experiment, simultaneous observation of all de-excitation paths is required. The combination of efficient detection of both $\gamma$ rays and conversion electrons, as demonstrated by the {\sc sage} spectrometer \cite{Pakarinen2014}, can be considered as one of the latest milestones in in-beam spectroscopy. The advent of radioactive beam facilities, such as the recently commissioned HIE-ISOLDE post-accelerator at CERN \cite{Rodriguez2016FirstHIE-Isolde}, have allowed for multi-step Coulomb-excitation experiments employing radioactive ion beams to be performed. For the analysis of Coulomb-excitation data it is of particular importance to obtain complementary spectroscopic information, e.g. electron conversion coefficients \cite{Zielinska2016AnalysisCode}. In this paper a compact SPectrometer for Electron DEtection ({\sc spede}), which will be combined with the Miniball spectrometer \cite{Warr2013} at HIE-ISOLDE, is described. 

\section{The {\sc spede} conversion electron spectrometer}
\label{sec:description}
\begin{figure*}
\includegraphics[width=0.95\textwidth]{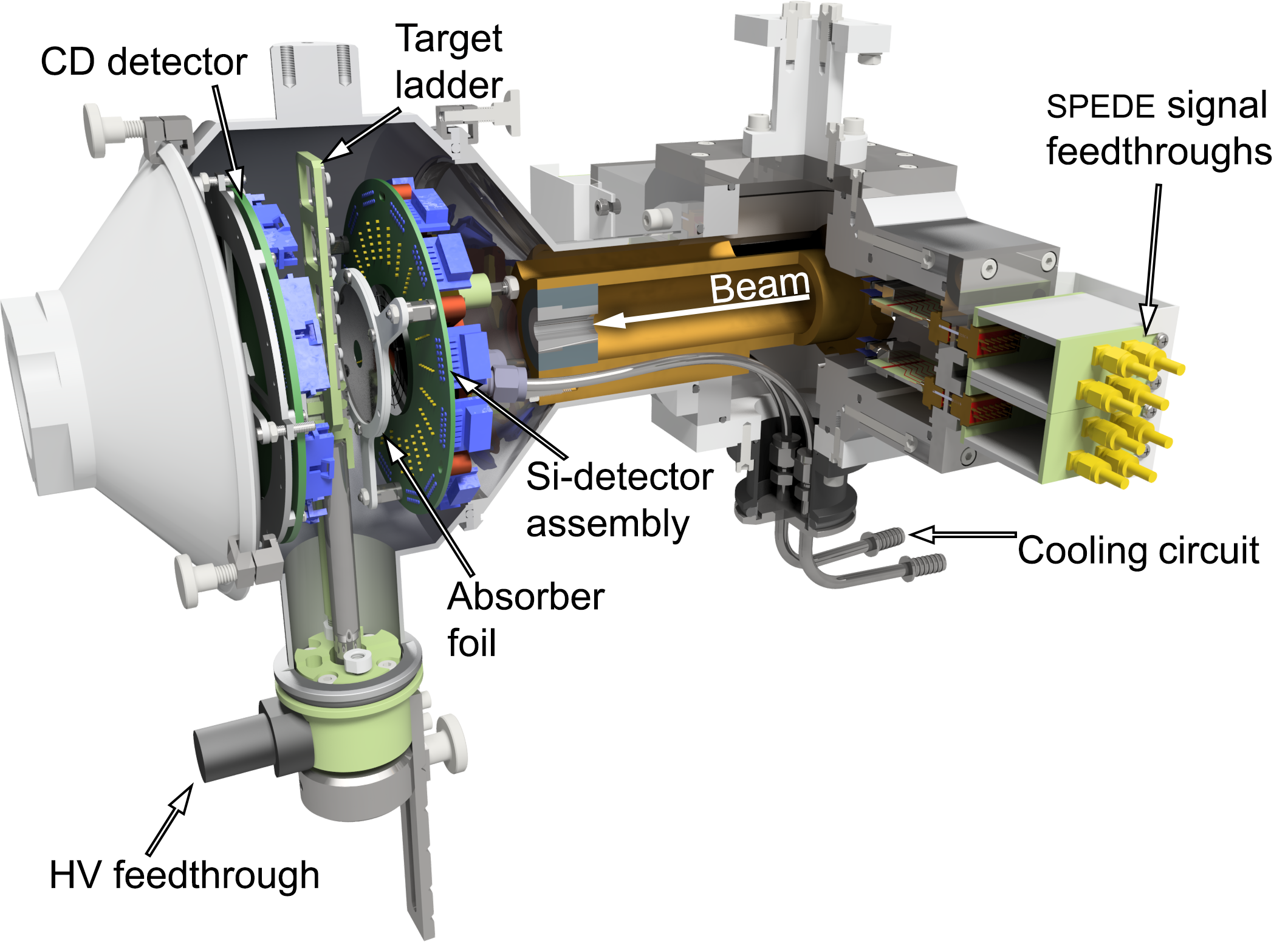}
\caption{{\sc spede} combined with the CD detector. Parts of the set-up are shown in cross-section for better visualisation. Essential components have been labelled and are discussed in more detail in the text. The white arrow indicates the beam direction.}
\label{fig:schematic}
\end{figure*}

The features of {\sc spede} and its compatibility with the Miniball spectrometer are described in the following subsections. Shown in Figure \ref{fig:schematic} is {\sc spede} including the Miniball particle detector for Coulomb-excitation experiments, a double-sided silicon strip detector (CD detector) \cite{Ostrowski2002}. The development of {\sc spede} has been discussed in detail in References~\cite{Konki2013,Papadakis2015,Cox2017}.

\subsection{Si-detector assembly}
\label{subsec:si-det}
{\sc spede} utilises a 24-fold segmented annular Si detector for the detection of internal conversion electrons. The detector is placed upstream and at a variable distance from the target. The conceptual design of {\sc spede} and the suppression of the $\delta$-electron background, discussed in more details in Subsection \ref{subsec:suppression}, assure that the conversion-electron signals are not overwhelmed by the high background. This allows for the placement of the detector in close proximity to the target, typically at a distance of 25\,mm, providing large angular coverage and detection efficiency of the order of 8$\%$ across a wide energy range. In Coulomb-excitation experiments the direct observation of electrons, i.e., without the use of magnetic fields for electron transportation, is essential for the kinematic correction of their velocities. To correct for the velocity, the angle between the detected electron and electron-emitting scattered particle is required. This can be obtained by using the {\sc spede} detector for the detection of electrons and the CD detector for the identification of scattered beam- and target-like particles, with the required angular, energy and time resolution.

\begin{figure}[h]
\includegraphics[width=0.49\textwidth]{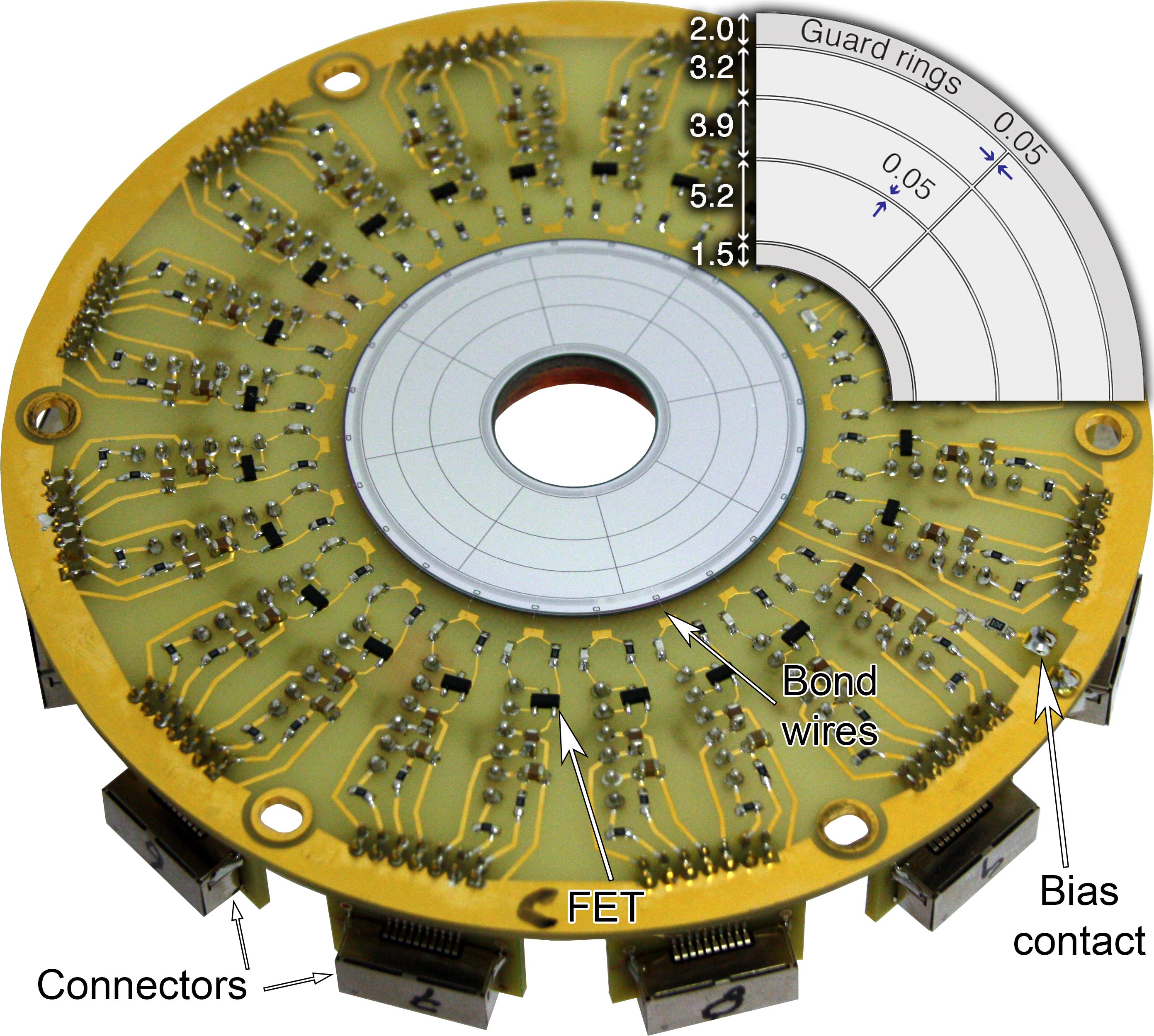}
\caption{Photograph of the {\sc spede} detector mounted on the PCB. The FETs and some of the filtering components can be seen around the detector. The dimensions of the individual detector segments are indicated in millimetres in the inset.}
\label{fig:detector}
\end{figure}

A 17\,mm aperture in the centre of the detector allows for the beam to pass through before impinging on the target. The 24 detector segments are arranged in three concentric rings around the beam aperture with 8 segments each. The segmentation scheme was optimised to provide sufficient position resolution for kinematic correction of the observed electron energies, while keeping the number of electronics channels relatively low \cite{Konki2013}. The widths of the segments (5.2, 3.9 and 3.2\,mm, from the innermost) were chosen so that their individual surface areas are kept approximately equal in order to keep the capacitance the same across all segments. The intersegment gaps are 0.05\,mm wide leading to a total detector inactive area of less than 2$\%$. The detector incorporates inner and outer guard rings to guarantee a homogeneous potential across the active area and minimise edge effects, also the bonding pads are distributed around the outer edge of the detector. A photograph of the detector mounted on the Printed Circuit Board (PCB) is presented in Figure~\ref{fig:detector} together with the detector geometry. The thickness of the detector presented here is 500\,$\mu$m and is optimised for electron energies up to 400\,keV. Detectors with a thickness of up to 1500\,$\mu$m can be produced.

\begin{figure}[h]
\includegraphics[width=0.49\textwidth]{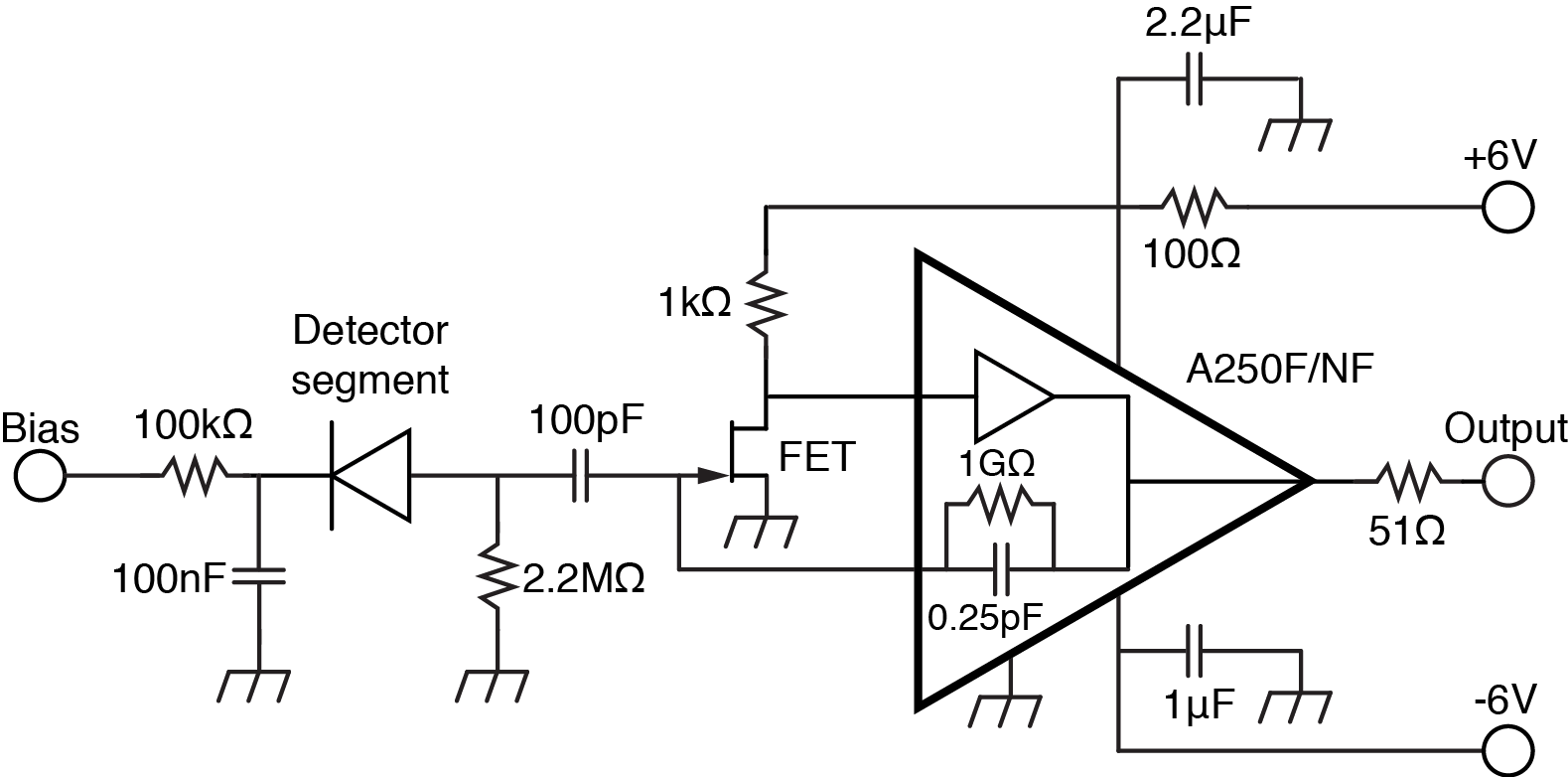}
\caption{Circuit diagram of a single {\sc spede} electronics channel. The bias and preamplifier power supply filtering circuits can be seen in the diagram together with the feedback characteristics of the preamplifier.}
\label{fig:circuit}
\end{figure}

Each detector segment is connected to an AMPTEK A250F/NF charge-sensitive preamplifier through a \linebreak PMBF4393 surface mount field effect transistor (FET). The gain of the preamplifiers is 4\,V/pC, equivalent to 175\,mV/MeV. Both the FETs and preamplifiers are positioned on the detector PCB inside the vacuum chamber. The preamplifier requires $\pm$6\,V, which is filtered by dedicated circuits on the PCB. The bias voltage is supplied on the rear common cathode of the detector through a 100\,k$\Omega$ bias resistor, which is decoupled to ground through a 100\,nF capacitor. The bias voltage for a detector of thickness 500\,$\mu$m is 90\,V and the leakage current at the operating temperature is less than 1\,nA. The preamplifiers are AC coupled to the detector segments through 100\,pF capacitors. A 2.2\,M$\Omega$ resistor is connected to each individual segment to act as the return path for the current from the detector. Due to the output drive limitations of the preamplifiers a 51\,$\Omega$ resistor was added in series with each output to prevent preamplifier loading and to back terminate the 50\,$\Omega$ cable impedance. The circuit diagram for a single detector segment is presented in Figure~\ref{fig:circuit}.

\begin{figure}[h]
\includegraphics[width=0.49\textwidth]{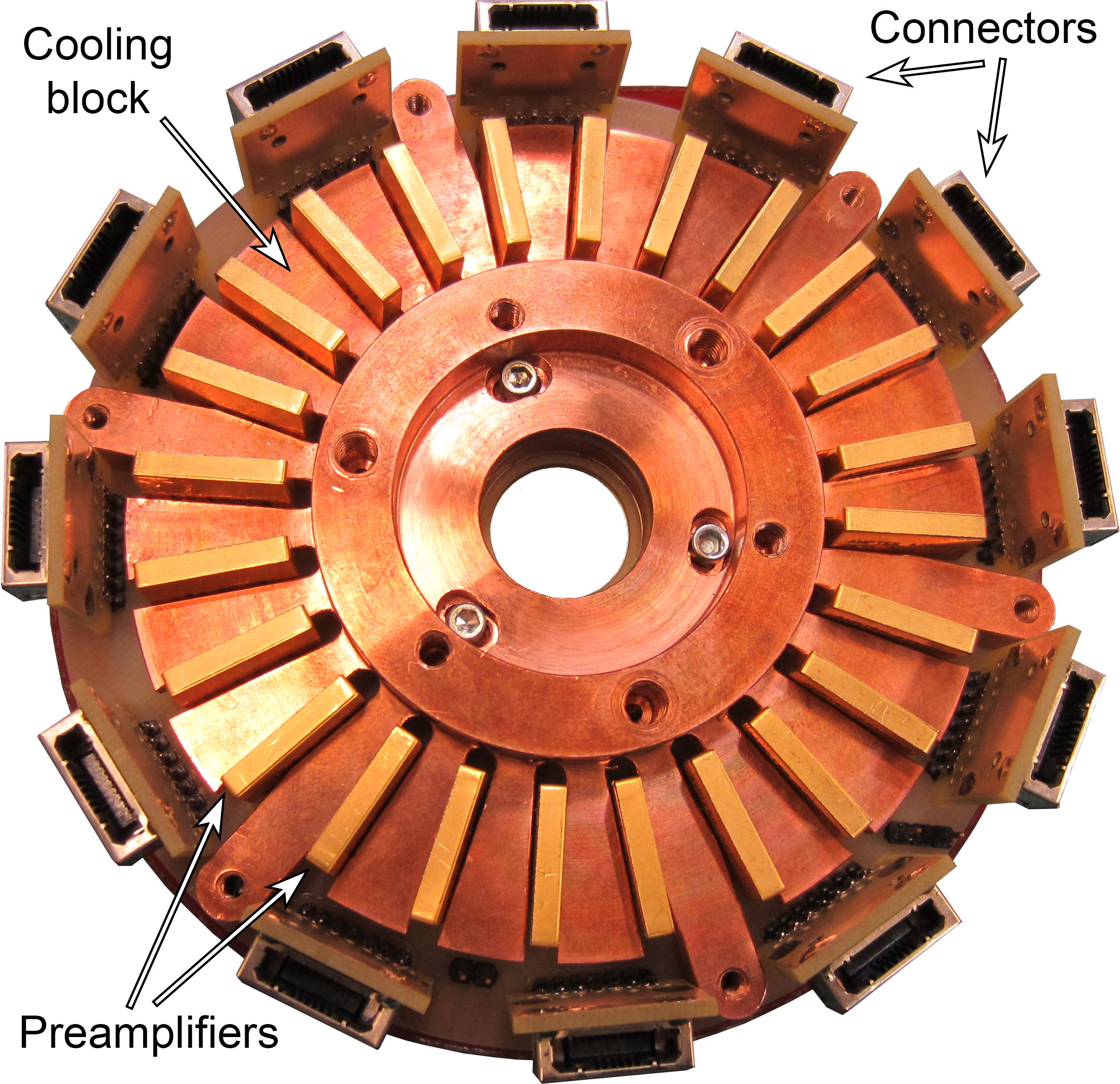}
\caption{Photograph of the rear of the {\sc spede} detector PCB mounted on the cooling structure. The preamplifiers can be seen between the segments of the fan-like cooling block.}
\label{fig:cooling}
\end{figure}

Two electronics channels are attached to each connector shown in Figures~\ref{fig:detector} and \ref{fig:cooling}. Micro-coaxial cable assemblies transport the signal from the connectors to the signal feedthroughs shown in Figure \ref{fig:schematic}. These assemblies also contain the preamplifier power cables.

To increase the stability of the Si detector and reduce electronic noise, the detector and preamplifiers are cooled. The detector PCB is mounted on a copper block housing an ethanol cooling circuit, that also integrates a fan-like structure surrounding the individual preamplifiers. The ethanol is circulated by a Julabo CF40 cryo-compact circulator with factory specified temperature range between -40$^\circ$C and +150$^\circ$C and cooling capacity of 0.12\,kW at -30$^\circ$C. During normal operating conditions the detector temperature is in the region of -5$^\circ$C to 0$^\circ$C.

\subsection{Suppression of electron background}
\label{subsec:suppression}
Conventionally, electron spectrometers have employed magnetic fields to transport electrons away from the vicinity of the production target. Consequently, magnetic lenses \cite{Kantele1975ASelection} or high-voltage barriers \cite{Butler1996} have been re becausequired to suppress the overwhelming $\delta$-electron background produced in the collisions between fast, heavy ions and target atoms. The fact that {\sc spede} is operated with low-intensity radioactive ion beams already reduces the $\delta$-electron rate compared to in-beam experiments with stable-ion beams of higher intensity. Segmenting the detector into 24 pixels further reduces the $\delta$-electron rate per channel. In addition, the following measures have been utilised to reduce the $\delta$-electron background in {\sc spede}.

\subsubsection{Detection geometry}
\label{subsubsec:geometry}
The {\sc spede} detector is positioned upstream of the target because $\delta$ electrons are primarily forward focused and the flux at backwards angles is lower by several orders of magnitude \cite{Bechthold1997}. An additional benefit of this detection geometry is the reduction of kinematic broadening, which is lowest when the trajectories of the electron-emitting particle and the electron are collinear.

\subsubsection{Introduction of high voltage on the target}
\label{subsubsec:hv}
{\sc spede} utilises a target ladder with four target positions, interchangeable from outside of the vacuum chamber. The majority of the $\delta$-electron flux is concentrated below 5\,keV \cite{Bechthold1997}. For this reason high voltages up to +5\,kV may be applied to the target in use. The high voltage decelerates the emitted electrons and reduces the number of low-energy $\delta$ electrons reaching the {\sc spede} detector. To ensure application of the voltage only on the target in use, the target ladder is made of insulating PEEK plastic, with metal inserts for mounting the targets and applying voltage. Contact is made through a spring arm with the in-beam target mounting frame, thus ensuring that the target integrity is not compromised. The remaining target frames may either be left floating or grounded through similar spring arm contacts.

\begin{figure}
\includegraphics[width=0.49\textwidth]{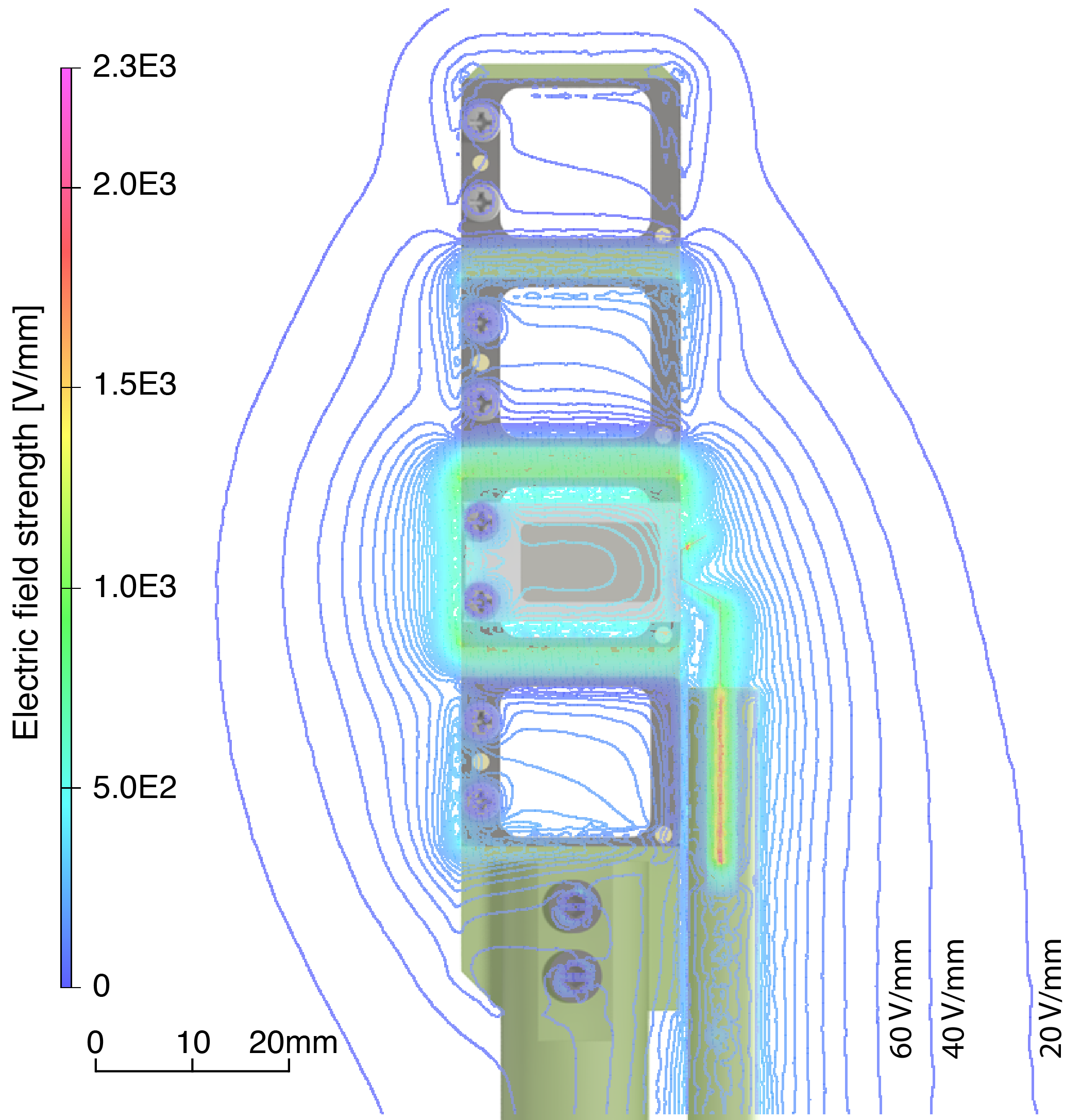}
\caption{Target ladder structure overlaid with the electric field lines in the target plane. In this simulation +5\,kV is applied on the in-beam target while the other target frames are left floating. The first three of the equipotential lines are marked with their respective values.}
\label{fig:Efield}
\end{figure}

The electric field profile at the target was simulated using the {\sc opera 3d} simulation package \cite{Fields}. These simulations were used in designing the target ladder to avoid regions of high electric field density which might lead to high-voltage discharges. All components in close proximity to the target were included in the simulation. Figure~\ref{fig:Efield} shows the target ladder geometry overlaid with the simulated electric field strength. The electric field profile was also included in the {\sc Geant4} \cite{Agostinelli2003} simulations presented later in this paper.

\subsubsection{Absorber foil}
\label{subsubsec:foil}
An aluminised Mylar foil is positioned between the target and the {\sc spede} detector to absorb low-energy electrons which escape the target. Any material in the path of the electrons affects the efficiency and effective resolution of the detector, thus, the absorber foil thickness is chosen to balance requirements of both suppression and resolution. The results presented in this paper were obtained using a 12\,$\mu$m (1.68\,mg/cm$^2$) Mylar foil with 0.5\,$\mu$m (0.14\,mg/cm$^2$) aluminium deposited on the surface. 

In order to obtain additional shielding against external noise, the aluminised Mylar foil is kept at ground potential. Moreover, the opaque nature of the foil in combination with a plastic ring covering the area between the detector and foil acts as a shield against the possible fluorescence light emitted from certain target materials.

\subsubsection{Suppression of random $\beta$-decay background}
\label{subsubsec:random}
Background of random $\beta$-decay origin will be suppressed by requiring a coincidence between the detected electrons in the {\sc spede} detector and the scattered particle observed in the CD detector. After a coincidence condition is applied, $\beta$ particles will be typically included only as random correlations, thus the background of $\beta$-decay origin should be suppressed significantly.

\subsubsection{Effect of the high voltage and absorber foil on spectral quality}
\label{subsubsec:effects}

\begin{figure}[!h]
\includegraphics[width=0.49\textwidth]{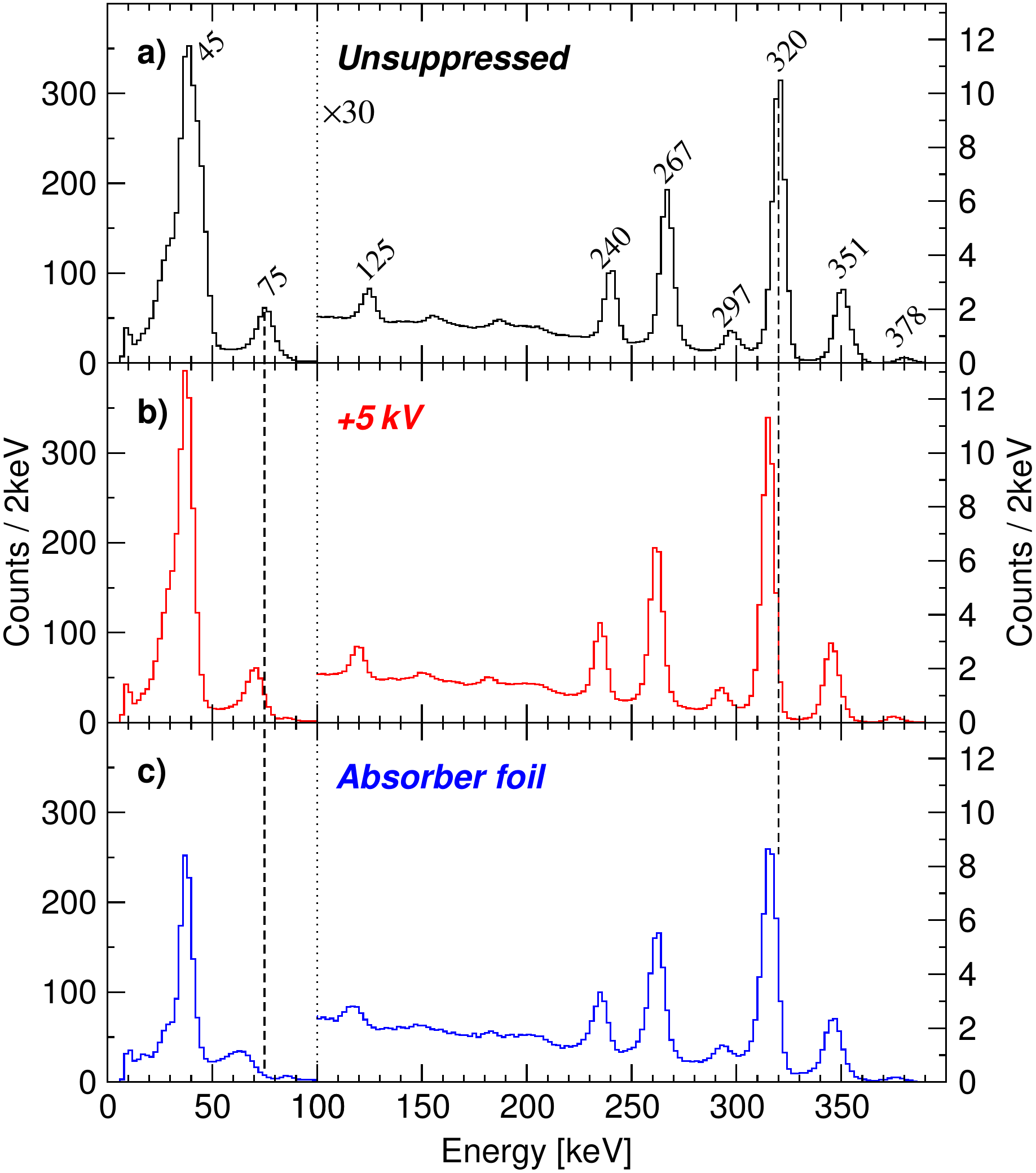}
\caption{
Electron energy spectra for $^{133}$Ba measured with {\sc spede} in various configurations. For energies higher than 100\,keV the y-axis has been expanded by a factor of 30 for visualisation purposes. The energy spectrum without suppression is shown in panel a). Panel b) shows the energy spectrum while applying a voltage of +5\,kV on the target and without absorber foil. Panel c) shows the energy spectrum measured with an aluminised Mylar absorber foil in front of the detector (without voltage on the target). The most prominent electron lines are labelled according to the corresponding energy.
}
\label{fig:133Ba_spectra}
\end{figure}

The effects of the high voltage and the absorber foil are shown individually in Figure~\ref{fig:133Ba_spectra}. A -5\,keV shift in the electron peak positions and no obvious change to the electron resolution are evident in the spectrum with high voltage. In the spectrum with absorber foil the electrons are shifted towards lower energies and the resolution is degraded due to straggling in the foil. This is more prominent for the lower-energy electrons. Indicative values of the effects of the high voltage and absorber foil at electron energies of 75\,keV and 320\,keV are presented in Table~\ref{tab:effects}. 

\begin{table}[h!]
\caption{Individual effects of the high voltage and aluminised Mylar absorber foil on the electron energy and full-width at half-maximum (FWHM) of the peaks in $^{133}$Ba \cite{Khazov2009}. 
}
\label{tab:effects}
\begin{tabular}{lllll}
\hline\noalign{\smallskip}
	 				& Energy	& FWHM		& Energy	& FWHM\\
					& [keV]		& [keV]		& [keV]		& [keV]\\
\noalign{\smallskip}\hline\noalign{\smallskip}
Literature			& 75.28(1)	&			& 320.03(1)	& \\
\textit{This work}	& 			& 			& 			& \\
Unsuppressed 		& 75.4(1)	& 9.7(1)	& 320.2(1)	& 6.6(1)\\
+5\,kV 				& 70.0(1)	& 9.6(1)	& 315.2(1)	& 6.9(1)\\
Absorber foil 		& 63.6(1)	& 12.9(1)	& 316.0(1)	& 7.7(1)\\
\noalign{\smallskip}\hline
\end{tabular}
\end{table}

The effect of the background suppression methods on the peak-to-total value (P/T) was investigated using a $^{133}$Ba source with a low-energy cut at 50\,keV. The spectrum with +5\,kV high voltage is of similar quality to the unsuppressed (P/T of 0.58(5)), whereas the spectral quality is reduced for the case where the aluminised Mylar foil is used (P/T of 0.41(5)).

\subsection{Electron detection efficiency}
\label{subsec:efficiency}
\begin{figure}
\includegraphics[width=0.49\textwidth]{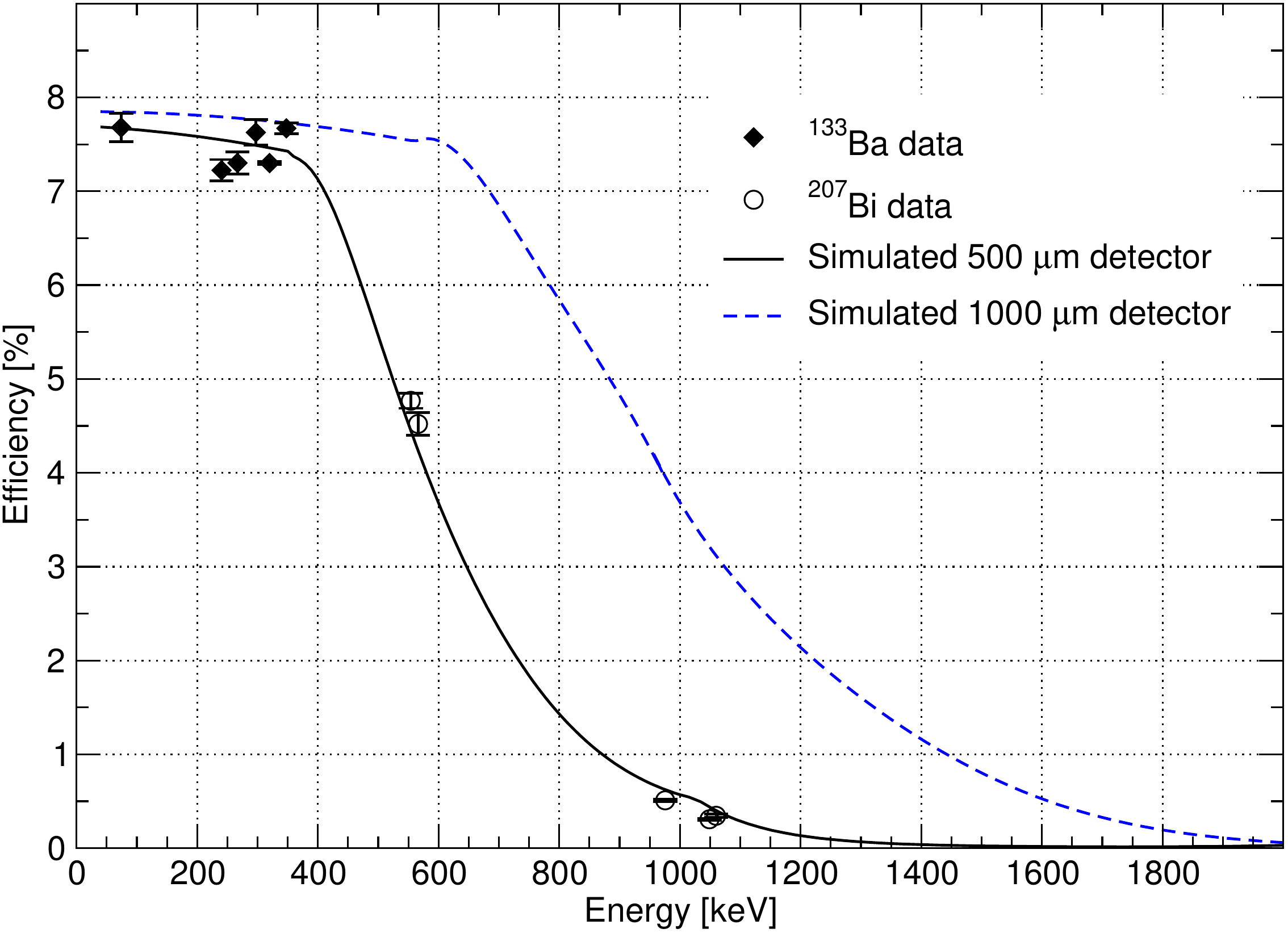}
\caption{Measured relative and simulated absolute detection efficiencies of {\sc spede} for a 500\,$\mu$m-thick detector. Filled diamonds and open circles denote data points obtained with $^{133}$Ba (normalised to the 320\,keV peak) and $^{207}$Bi (normalised to the 482\,keV peak) calibration sources, respectively. The simulated efficiency for a detector of thickness 1000\,$\mu$m is shown for comparison.}
\label{fig:SimEff}
\end{figure}

The relative detection efficiency of {\sc spede} was determined using open $^{133}$Ba and $^{207}$Bi calibration sources. The relative efficiency values were normalised to a simulated efficiency curve produced using {\sc Geant}4. The measured efficiency and simulated efficiency curve are presented in Figure~\ref{fig:SimEff}. The sources were mounted on the target ladder and no voltage or absorber foil were used. The simulation package has been developed within the {\sc nptool} framework \cite{Matta2016} and will be presented in more detail elsewhere \cite{Cox}. It contains the full Miniball spectrometer, the complete geometry of {\sc spede}, including the support structures, absorber foil and electric field produced by the high voltage applied to the target. To allow for comparison between the different set-ups the Miniball Coulomb-excitation target chamber\cite{Warr2013} has also been included.

The detection efficiency remains relatively constant for energies up to 400\,keV and close to the expected value calculated from geometrical constraints. At higher energies the detection efficiency decreases as the number of punch-through events and electrons not depositing their full energy in a single segment increases. For detection of higher-energy electrons a thicker detector could be used. For comparison the simulated efficiency curve with a 1000\,$\mu$m detector is also plotted in Figure~\ref{fig:SimEff}. The efficiency in the region of 1\,MeV is increased by a factor of 7, while the efficiency below 400\,keV remains largely unaffected. The aluminised Mylar absorber foil used for the results presented in this paper reduces the efficiency below $\sim$150\,keV, but does not have a significant effect at higher electron energies.

\subsection{Combining with Miniball}
\label{subsec:combining}
\begin{figure}
\includegraphics[width=0.49\textwidth]{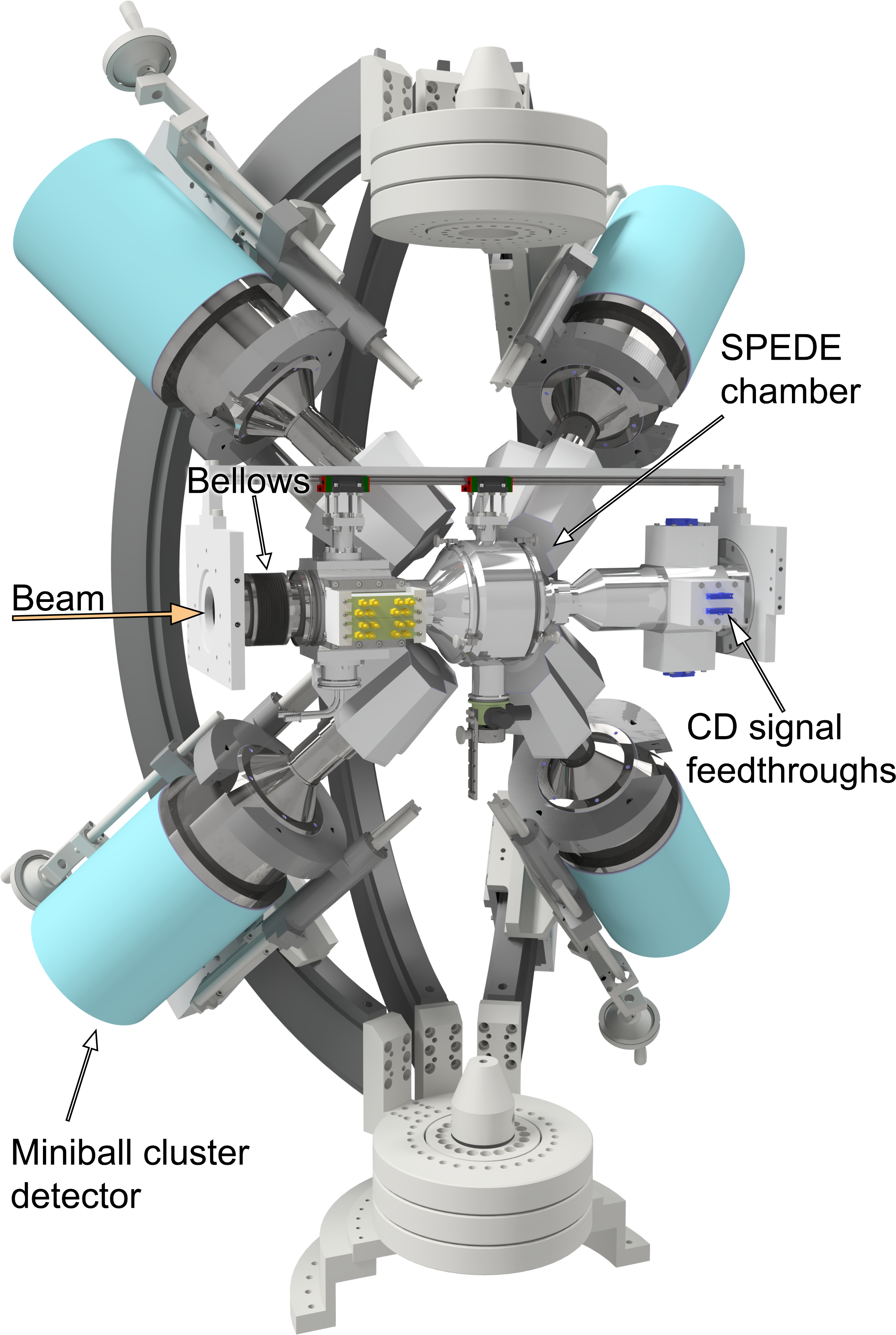}
\caption{{\sc spede} combined with the Miniball spectrometer. The target chamber support tables and one hemisphere of the Ge-detector array have been removed for better visualisation. Main parts and the beam from HIE-ISOLDE are labelled.}
\label{fig:miniballschematic}
\end{figure}

{\sc spede} was designed to be compatible with the existing Miniball infrastructure at HIE-ISOLDE, CERN. {\sc spede} in the centre of Miniball spectrometer is shown in Figure~\ref{fig:miniballschematic}.

A compact aluminium target chamber with 2.5\,mm thick walls houses the {\sc spede} detector, target ladder and the CD detector. The beam pipe and CD-detector \linebreak feedthroughs of the Miniball Coulomb-excitation target chamber are connected to the downstream part of the {\sc spede} chamber, while the CD detector is mounted on a purpose-built support. A beam pipe accommodating the {\sc spede} detector feedthroughs is connected upstream of the target chamber. This beam pipe includes the support structure for the {\sc spede} detector, the beam collimator and the ethanol cooling circuit pipes (see Figure~\ref{fig:schematic}).

During operation the {\sc spede} target ladder allows the use of any one of the four targets without opening the target chamber. Access to the {\sc spede} or CD detector can be gained by removing the bellows (see Figure~\ref{fig:miniballschematic}) and sliding the target chamber on a rail system in the upstream direction. 

The design of the Miniball frame enables the placement of the cluster Ge detectors at various angles and rotations. This allows for the use of different target chambers and ancillary detectors without compromising the $\gamma$-ray detection efficiency.

\section{Performance of {\sc spede}}

\subsection{Simultaneous $\gamma$-ray and conversion-electron spectroscopy at HIE-ISOLDE}
\label{subsec:coincidences}
\begin{figure}[!ht]
\includegraphics[width=0.49\textwidth]{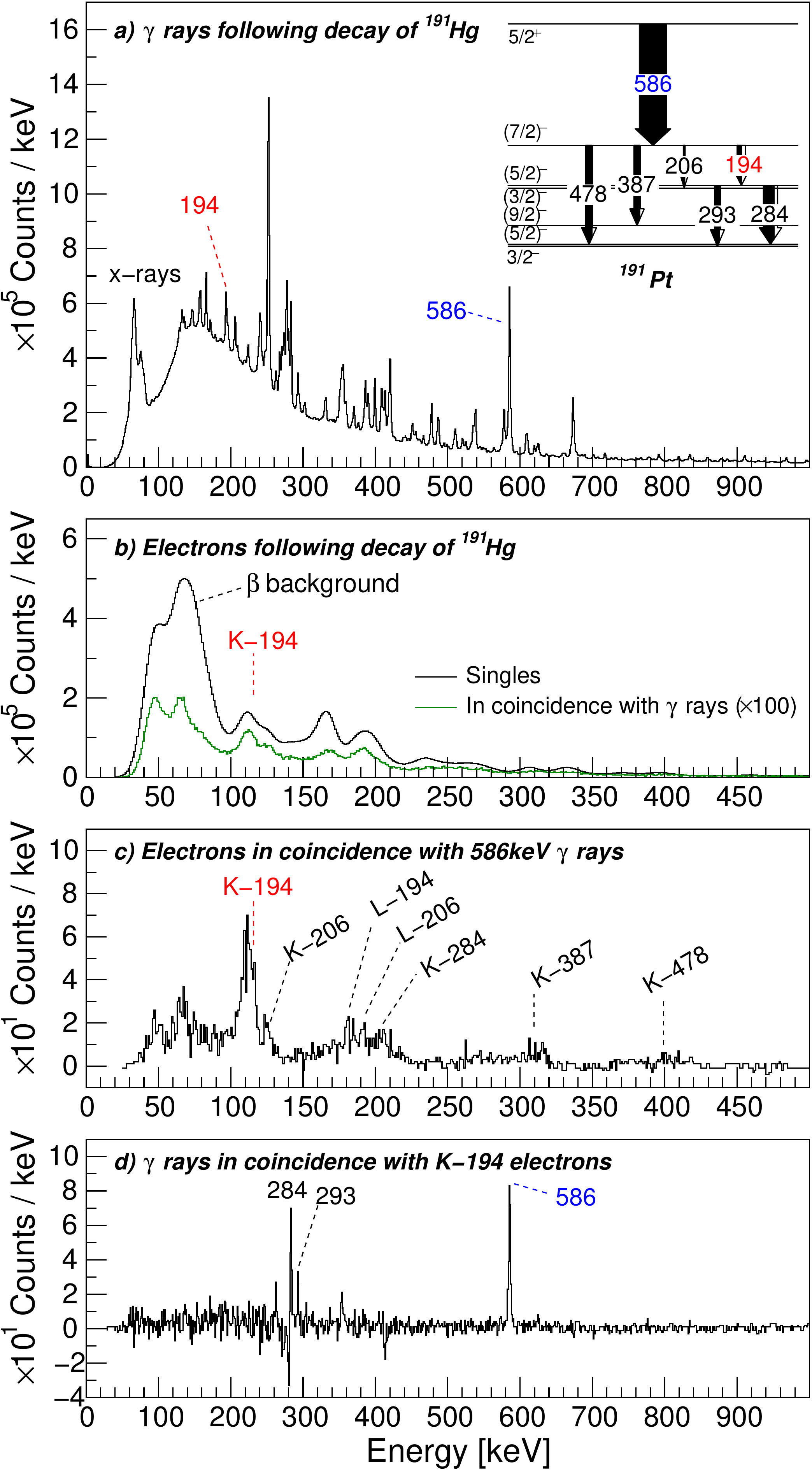}
\caption{Simultaneously measured $\gamma$-ray and conversion-electron energy spectra following the decay of $^{191}$Hg nuclei. In panels a) and b), the total singles $\gamma$-ray and electron energy spectra, respectively, are shown. The green curve in panel b) corresponds to the electrons observed in coincidence with any $\gamma$ ray. c) Background-subtracted electron energy spectrum in coincidence with the 586\,keV $\gamma$ rays. d) Background-subtracted $\gamma$-ray energy spectrum in coincidence with the 116\,keV electrons (K-electron line of the 194\,keV transition in $^{191}$Pt). A partial level scheme of $^{191}$Pt is shown in the inset \cite{Johansson1978}. The width of the arrows are proportional to transition intensities (black and white components correspond to the $\gamma$-ray and conversion-electron intensities, respectively). Relevant $\gamma$-ray and conversion-electron lines are labelled.}
\label{fig:coincidences}
\end{figure}

The integration of {\sc spede} with the Miniball spectrometer was tested using an implanted radioactive $^{191}$Hg source. The decay chain $^{191}$Hg($t_{1/2}=$ 50\,min) $\rightarrow$ $^{191}$Au(2.2\,h) $\rightarrow$ $^{191}$Pt (2.8\,d) provided a high-statistics data set to perform simultaneous $\gamma$-ray and conversion-electron ($\gamma$e$^\textrm{-}$)\linebreak spectroscopy. This is illustrated in panel a) of Figure~\ref{fig:coincidences}, where the total singles $\gamma$-ray energy spectrum obtained with the aforementioned source is shown. Panel b) of Figure~\ref{fig:coincidences} shows the total singles electron energy spectrum and electrons in coincidence with any $\gamma$ ray. It should be noted that the electron energy spectra include background arising from the $\beta^+$ decay of $^{191}$Hg \cite{Baglin2009Nuclear179}. The $\gamma$e$^\textrm{-}$ performance is demonstrated in Figures~\ref{fig:coincidences} c) and d), where gates on the $\gamma$e$^\textrm{-}$ matrix are set on the 586\,keV $\gamma$ ray and the 116\,keV electron lines (corresponding to the K-electron line of the 194\,keV transition in $^{191}$Pt), respectively.

\subsection{In-beam testing of {\sc spede} at JYFL-ACCLAB}
\label{subsec:inbeam}
The in-beam performance of {\sc spede} has been investigated in a Coulomb-excitation experiment employing the\linebreak$^{82}$Kr($^{197}$Au,$^{197}$Au*) reaction at 4.26\,MeV/u beam energy in the Accelerator Laboratory of the University of\linebreak Jyv\"askyl\"a (JYFL-ACCLAB), first reported in \cite{Cox2017}. The target was a self-supporting 1200\,$\mu$g/cm$^2$ thick gold foil. The beam from the K130 cyclotron was chopped to provide bunches of 200\,$\mu$s beam-on and 800\,$\mu$s beam-off to replicate the typical time structure of the HIE-ISOLDE beam. The performance of {\sc spede} was tested up to $50\times 10^6$\,pps impinging on the target. In typical running conditions the beam intensity was $\sim$$2\times 10^6$\,pps, resulting in an average counting rate of 1300\,Hz/channel in the {\sc spede} detector. Scattered particles were detected with an array of six 1\,cm$^2$ PIN diodes described in \cite{Cox2017}.

\begin{figure}[h]
\includegraphics[width=0.49\textwidth]{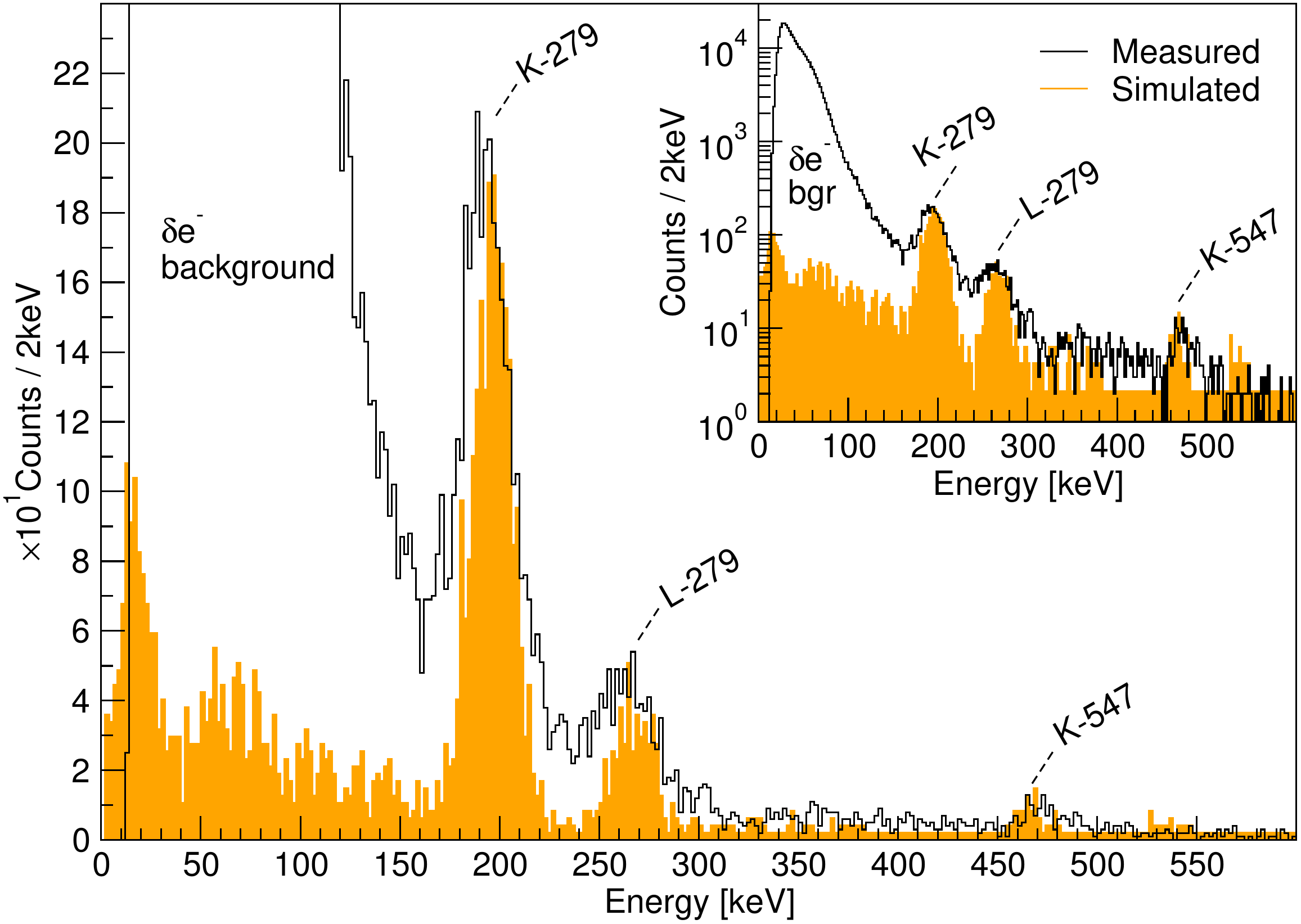}
\caption{Measured (black) and simulated (orange) particle-gated electron energy spectra, kinematically corrected for $^{197}$Au. The K- and L-conversion electron lines from the 279\,keV transition and the K-conversion line from the 547\,keV transition in $^{197}$Au are labelled. Spectra with full-range logarithmic y axis are shown in the inset.}
\label{fig:electronSpectrum1}
\end{figure}

Figure~\ref{fig:electronSpectrum1} shows the measured and simulated particle-gated electron energy spectra. Both spectra are kinematically corrected for $^{197}$Au and were obtained using the aluminised Mylar absorber foil and no high voltage on the target. The simulated spectrum reproduces the main structures in the measured spectrum and the ratio between the detected peaks. The complexity of the measured background makes its accurate simulation unrealistic and computationally demanding. 

A K/L conversion ratio of 5.7(7) for the 279\,keV \linebreak$\frac{5}{2}^{+}\rightarrow\frac{3}{2}^{+}$ transition in $^{197}$Au was extracted from the data shown in Figure~\ref{fig:electronSpectrum1} and using the efficiency presented in Figure~\ref{fig:SimEff}. The value obtained is compared with the K/L ratios calculated using the BrIcc conversion coefficient calculator \cite{Kibedi2008} for pure M1 and E2 transitions and with a mixed M1+E2 transition using the mixing ratio of \linebreak$\delta$=-0.39(2) \cite{Stuchbery1988Measured197Au}. The measured value is in excellent agreement with the calculated value for a mixed transition.

\begin{table}[h]
\caption{Comparison of the K/L conversion ratio for the 279\,keV $\frac{5}{2}^{+} \rightarrow \frac{3}{2}^{+}$ transition in $^{197}$Au obtained with {\sc spede} to values calculated with BrIcc \cite{Kibedi2008} and literature \cite{Stuchbery1988Measured197Au}.
}
\label{tab:K/L}
\begin{tabular}{lllll}
\hline\noalign{\smallskip}
			& M1		& E2		& M1+E2 		&  \\
            & (BrICC)  	& (BrICC)	& (literature)	& This work \\
\noalign{\smallskip}\hline\noalign{\smallskip}
K/L ratio	& 6.05(12)	& 1.85(4)	& 5.63(13)		& 5.7(7)\\
\noalign{\smallskip}\hline
\end{tabular}
\end{table}

\section{Summary}
\label{sec:summary}
{\sc spede} has been built and commissioned at JYFL-ACCLAB and at the HIE-ISOLDE facility, CERN. {\sc spede} combines a segmented Si detector for the measurement of internal conversion electrons directly from the target with the Miniball spectrometer for the measurement of $\gamma$ rays and scattered particles. The off-line tests indicate that the spectrometer works within the design criteria with a detection efficiency of the order of 8$\%$ and FWHM at 320\,keV in the region of 6-8\,keV depending on the running conditions. The first in-beam tests have shown the power of the spectrometer for the direct detection of conversion electrons and the extraction of conversion coefficients. A comprehensive simulation package for the set-up was developed using the {\sc nptool} framework in {\sc Geant4} and can be used to investigate the feasibility of planned experiments.

We would like to acknowledge Magdalena Zieli\'nska Jose Alberto Rodriguez, Miguel Lozano Benito and Karl Johnston for their help and support during the tests at ISOLDE. The research leading to these results has received funding from the People Programme (Marie Curie Actions) of the European Union's Seventh Framework Programme (FP7/2007-2013) under REA grant agreement n$^\circ$ 304033, by the Academy of Finland (contract number 265023) and by the United Kingdom Science and Technology Facilities Council (Grant Ref. ST/J000094/1). The Miniball collaboration is acknowledged for supporting this project. In addition, the German BMBF under contracts 05P15RDCIA and 05P15PKCIA and "Verbundprojekt \linebreak05P2015" are acknowledged.

\bibliographystyle{epj}
\bibliography{Spede1.bib}

\end{document}